# Visualizing femtosecond dynamics with ultrafast electron probes through terahertz compression and time-stamping

Mohamed A. K. Othman[*], Emma C. Snively, Annika E. Gabriel, Michael E. Kozina, Xiaozhe Shen, Fuhao Ji, Samantha Lewis, Stephen Weathersby, Praful Vasireddy, Duan Luo, Xijie Wang, Matthias C. Hoffmann[†] and Emilio A. Nanni[††]

*SLAC National Accelerator Laboratory, Stanford University, Menlo Park, CA, 94025 USA*

[*]mothman@slac.stanford.edu, [†]hoffmann@slac.stanford.edu, [††]nanni@slac.stanford.edu

**Visualizing ultrafast dynamics at the atomic scale requires time-resolved pump-probe characterization with femtosecond temporal resolution. For single-shot ultrafast electron diffraction (UED) with fully relativistic electron bunch probes, existing techniques are limited by the achievable electron probe bunch length, charge, and timing jitter.[1–5] We present the first experimental demonstration of pump-probe UED with THz-driven compression and time-stamping that enable UED probes with unprecedented temporal resolution. This technique utilizes two counter-propagating quasi-single-cycle THz pulses generated from two OH-1 organic crystals[6,7] coupled into an optimized THz compressor structure. Ultrafast dynamics of photoexcited bismuth films show an improved temporal resolution from 178 fs down to 85 fs when the THz-compressed UED probes are used with no time-stamping correction. Furthermore, we use a novel time-stamping technique to reveal transient oscillations in the dynamical response of THz-excited single-crystal gold films previously inaccessible by standard UED, achieving a time-stamped temporal resolution down to 5 fs.**

Ultrafast electron bunch probes with femtosecond-scale temporal duration provide a critical tool for interrogating the atomic and molecular structure of materials to reveal new dynamics.[1,2,8–12] Recently, UED facilities have provided a unique insight into visualizing elusive ultrafast processes from photochemical reactions and lattice motion,[13] to phase transitions[6] occurring in quantum materials.[14] The electron probes in these instruments are shaped by an ultrashort laser pulse irradiating a metal cathode, and then accelerated by a DC electric field[15] or radio-frequency linac.[2,4,5,9] The same laser system is also used as a pump to photoexcite a sample, providing cross timing synchronization between the laser pump and the electron probe. This approach allows for probing fast dynamics in thin films and two-dimensional matter, as well as the liquid and gas phases of materials.

However, in pump–probe experiments, the temporal resolution of measurements is not only limited to the probe bunch length but also suffers from timing jitter between the arrival times of the pump and probe pulses at the sample which obscures the fastest dynamical response time. Shot-to-shot jitter correction down to 10s of fs in XFELs has been reported using all-optical techniques.[16,17] For UED beamlines, rf-based ballistic velocity modulation of electrons has been shown to provide sub-10 fs UED probes [18,19] but it cannot correct time-of-arrival (TOA) jitter. Alternatively, the recent developments of a double bend achromatic beamline[4,5] demonstrated that a 30 fs temporal resolution can be attained in time-resolved UED experiments. However, such an approach is limited by the space-charge induced by energy chirp, requires significant infrastructure, and limits the charge and/or spatial resolution in order to achieve these levels of temporal resolution.

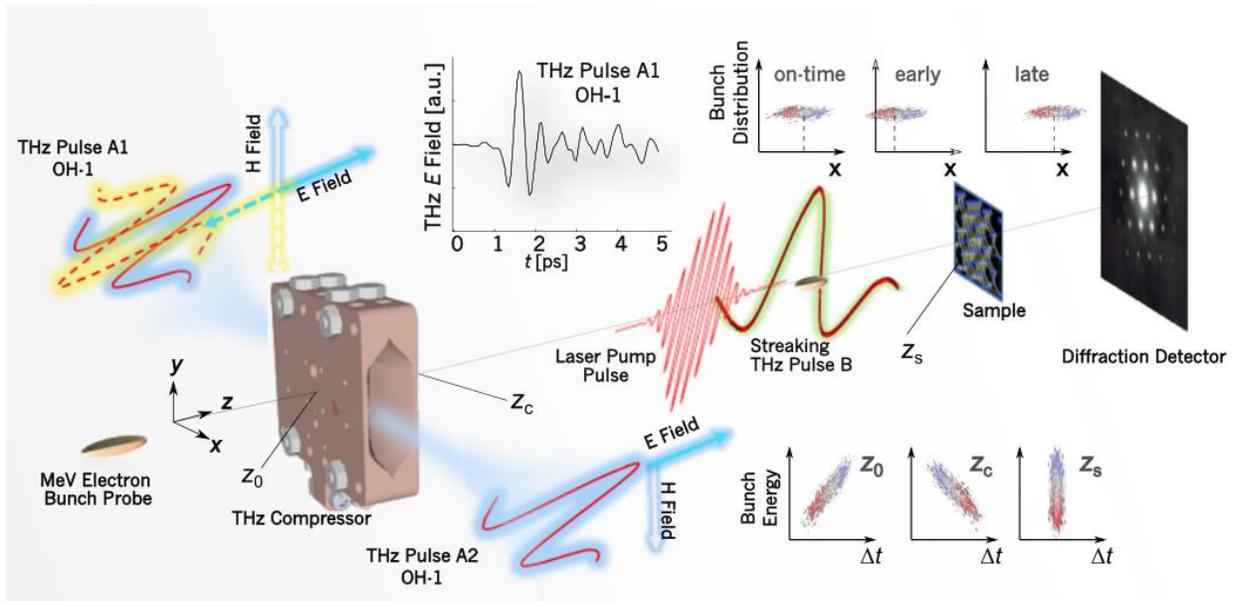

**Fig. 1 | Schematic of the SLAC's MeV-UED instrument with dual-feed THz time-stamping.** A 3 MeV, <10 fC electron probe interacts with two counter-propagating THz pulses causing probe length reduction at the sample position 1 m downstream. Transverse momentum due to THz pulse B streaks the longitudinal profile of the beam along the y axis on a downstream phosphor detector. THz pulses A1 and A2, obtained from optical rectification of a 1300 nm laser pulse in organic OH-1 crystals, are polarized along the z axis and provide up to 130 MV/m of peak electric field inside the THz compressor structure. THz pulse B is obtained from optical rectification in LN, polarized along the y axis, and used for streaking. A collinear 800 nm laser pump pulse is also incorporated into the sample chamber setup. The inset at the top of the schematic shows the measured THz field inside the copper THz compressor structure using electro-optic sampling of THz pulse A1 (spectrum is centered around 1.7 THz see Supplementary Section 2).

The advent of ultrafast THz technologies has introduced an array of instrumental capabilities for enhanced UED science, from beam manipulation at THz frequencies,[20–24] to pump-probe systems harnessing the THz frequency range to induce dynamical behavior in materials.[25–27] Efficient interaction of strong THz fields with electrons can offer higher energy modulation compared to conventional rf cavities. Ultra-bright electron probe beams with 10s of fs jitter[20,28–30] have also been demonstrated through dramatic THz-driven compression in both sub-relativistic and relativistic beam energy regimes.[28,31,32] In particular, these works have shown that small-footprint THz structures with strong field gradients could be used to produce compressed relativistic electron probes with minimal jitter, aided by the inherent synchronization between the THz pulse and the drive laser for the probe beam. In this article, we demonstrate unprecedented temporal resolution in UED experiments with laser-driven THz radiation for velocity bunching and time stamping. Using this novel technique, we visualize, for the first time, femtosecond dynamics of THz time-stamped electron probes in UED measurements in both single-crystal and polycrystalline thin films.

**CONFIGURATION OF MEV-UED WITH THZ-INDUCED COMPRESSION AND TIME-STAMPING**

We have modified the MeV-UED beamline at SLAC National Accelerator Laboratory to carry out pump-probe experiments with an electron probe that has been compressed and time-stamped using terahertz radiation. The beamline schematic is shown in Fig. 1 (more detailed schematic in Supplementary Fig S1). The 3 MeV, few fC electron probes were generated from a rf photocathode gun at a repetition rate of 360 Hz. The longitudinal compression and time-stamping of the relativistic electron probe was accomplished through interaction with two counter-propagating quasi-single-cycle THz pulses with a center frequency of 1.7 THz, generated via optical rectification of NIR (λ=1300 nm) laser pulses in two OH-1 organic crystals. The interaction between the THz pulses and electrons occurs in an optimized THz structure providing dispersion-free focusing below the diffraction limit of the THz fields and leading to strong field enhancement[22] (Supplementary Section 2). The electric field orientation of the THz pulse (Fig. 1) is selected by aligning the OH-1 crystallographic axis with the NIR laser polarization. We demonstrate two modes of operation: a parallel (P)-mode in which the two OH1 crystals have their crystallographic axes parallel to the pump laser polarization, and an antiparallel (AP)-mode in which one of the crystals is flipped by 180 degrees about the optical axis. The setup also includes a THz-driven streaking diagnostic[33] in which a THz-excited ~100 μm metallic slit is located one meter downstream from the compressor stage, to characterize the temporal properties of the electron probe (Supplementary Section 1).

At the optimal temporal overlap of the counter-propagating THz pulses with the electron probe, an energy modulation is induced as the electrons travel the 70 μm accelerating gap in the THz compressor structure. Ideally, the electron probe is injected into the THz structure when the superposition of the THz waveforms produces a zero-crossing with maximum electric field slope. The energy of the tail or late electrons increases relative to the head or early particles, providing an energy modulation of the electron probe up to ~10 kV/100 fs (Supplementary Section 2-3). As a result, the shot-to-shot beam longitudinal profile is compressed and timing jitter from time-of-flight variation is suppressed after a subsequent drift of one meter. Using General Particle Tracer (GPT)[34] simulations to model the effect of the THz compression with a full, 3D electromagnetic field distribution indicates that the peak field applied on the electrons from each arm is about 65 MV/m (Supplementary Section 3). For the P-mode, the THz-electron interaction in the compressor also leads to a strong transverse time-momentum correlation due to the presence of deflecting magnetic fields. In particular, the late portion of the electron probe is deflected along the positive x-direction, while the early portion is deflected toward the negative x-direction imparting a time-stamp on the probe's spatial profile at the detector.

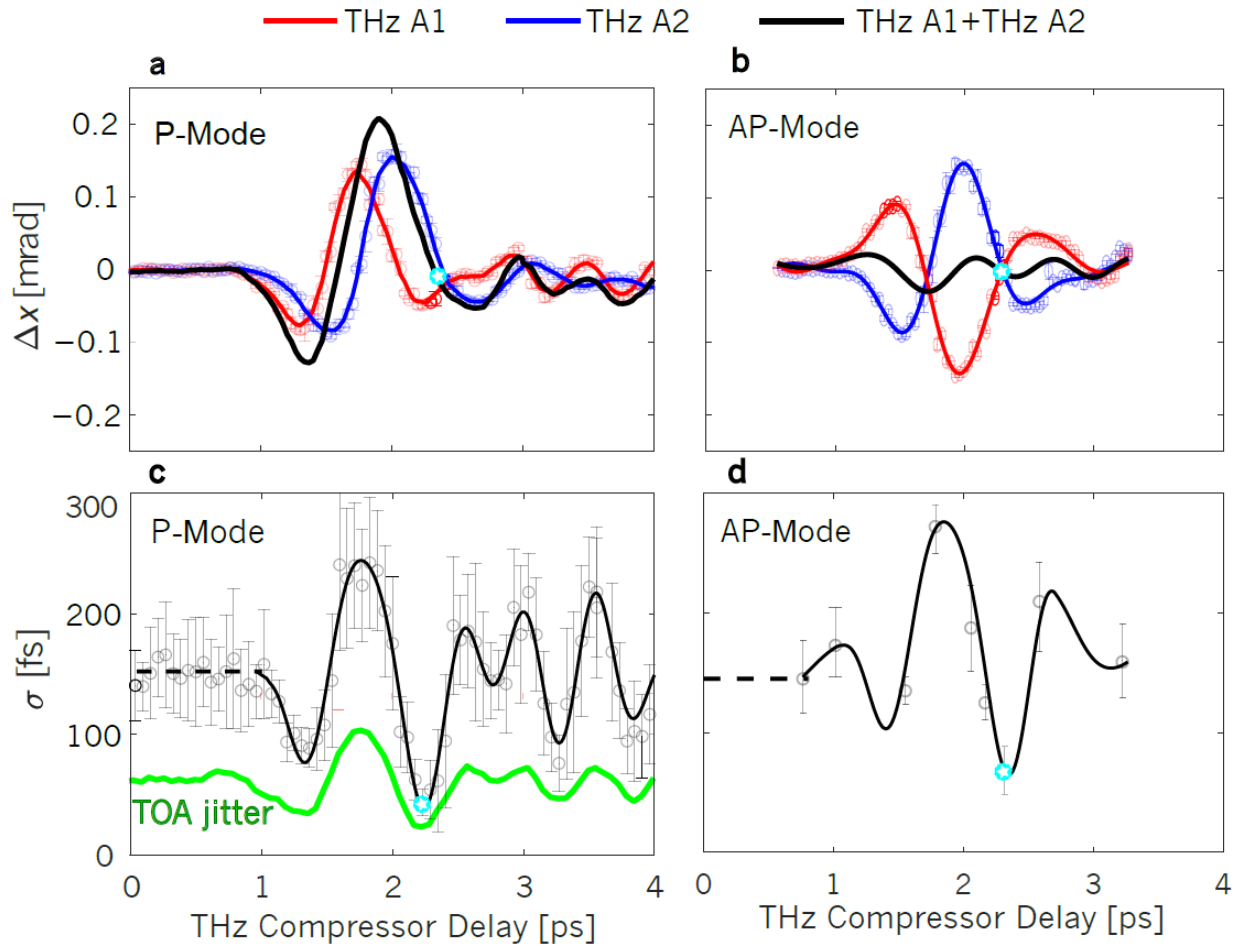

**Fig. 2 | Single-shot measurements of THz time-stamped electron probe dynamics.** The two modes of operation are shown, with the crystallographic axes of both crystals oriented in parallel, i.e., the P-mode (a,c) and anti-parallel, i.e., the AP-mode (b,d). (a-b) Time-dependent deflection Δx and (c-d) electron probe length σ for the P- and AP-modes are shown. The absolute minimum average probe length was found to be 40 fs r.m.s. and 69 fs r.m.s. for P and AP-mode respectively down from about 154 fs r.m.s. The TOA jitter is also shown for the P-mode in (c) indicating a minimum of 23 fs r.m.s. Note that the THz interaction does not produce a transverse deflection at the optimum phase for compression timing (at ~2.2 ps in a-c and ~2.3 ps in b-d as designated by the marker).

Fig 2(a-b) shows the deflection of electron probe centroid as the time of arrival is varied for the two THz pulses from both P and AP modes. Importantly, the transverse distribution can be correlated to the arrival time for time stamping, as demonstrated later. Both THz pulses induce similar deflection, indicating that the electron probe observes similar peak fields from both pulses as seen in Fig. 2(a-b). In the optimal compression phase, the electron probe centroid experiences no transverse deflection. For the P-mode, the two THz pulses were relatively delayed by about 200 fs to achieve the highest energy chirp resulting in compression by a factor of 3.85. A minimum electron probe temporal length is measured to be 40 fs r.m.s. down from an average of 154 fs r.m.s. in the delay scan in Fig. 3(c). A simultaneous improvement of the beam's shot-to-shot TOA is achieved, with a minimum TOA jitter of 23 fs r.m.s, down

from 69 fs r.m.s. This is evident from the reduced variation in the time-of-arrival measurement (Supplementary Fig. S4). For the AP-mode, the compression factor is 2.2, from 154 fs down to a minimum length of 69 fs r.m.s. Note that the temporal measurements of the uncompressed electron probe length have a larger spread of ±30 fs, compared to ±8 fs in the THz compression case, see Supplementary Section 4. The shot-to-shot stability of the compressed electron probe was analyzed in the P-mode over hundreds of subsequent single shots, and results are presented in Supplementary Fig. S5 and Section 5.

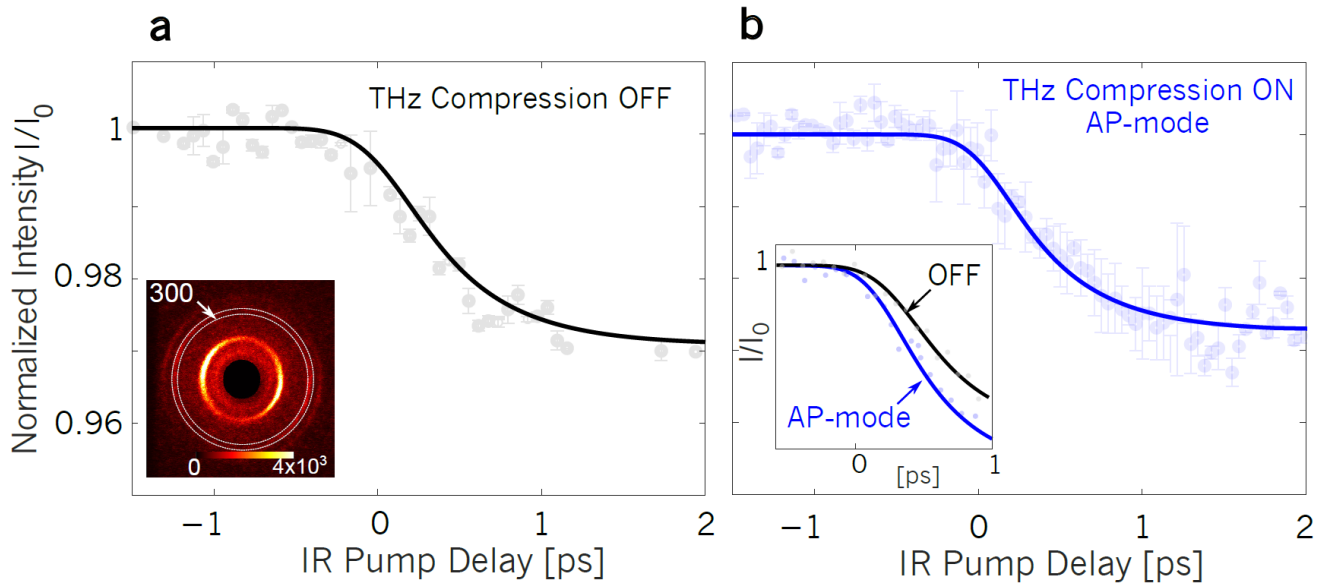

**Fig. 3 | Ultrafast Dynamics of the Bragg peak intensities demonstrating improved temporal resolution of UED in the AP-mode using an IR pump.** (a-b) Intensity of polycrystalline Bi (300) Bragg peaks after photoexcitation with an 800 nm pump laser, (a) without THz compression, and (b) with THz compression in the AP-mode. The overall temporal resolution is about (a) 178 fs and (b) 85 fs r.m.s. using the THz compressed UED probe in the AP-mode, showing an improvement of about a factor of 2.2. Diffraction patterns in the inset were taken by averaging 36 images.

**THE TEMPORAL RESOLUTION OF THE THZ COMPRESSED UED PROBE IN THIN BI FILMS**

We now demonstrate the enhanced temporal resolution provided by this THz-based compression and time-stamping method when applied to an electron probe characterizing ultrafast transient behavior in matter. The response time of the UED system can be obtained by measuring the switching time, i.e., how the diffraction signal changes with respect to a pump laser trigger. Extreme ultrafast beam manipulation can deteriorate the quality of the probes in terms of energy spread, transverse emittance, and pointing jitter, producing these collateral effects as the interaction compensates for the probe beam temporal distribution and jitter characteristics. In this experiment, the energy spread of the compressed electron probe increased by only 10% from the original probe, as observed at the sample, indicating that the measured $q$-space using UED is not significantly altered by the THz interaction. Our model also shows that the slice energy spread is reduced compared to our earlier work[28] due to the counter-propagating THz interaction symmetry in the compressor structure (see additional information in Supplementary Section 5). Using this operating configuration, the THz-compression and time-stamping process does not lead to a significant degradation in emittance. High quality diffraction patterns were obtained from the THz-compressed electron probe in both single crystals as well as polycrystalline thin films (Fig. 3a insert).

The improved temporal resolution enabled by the THz-driven compression was first characterized using time-resolved diffraction measurements on photoexcited bismuth (Bi) thin films to visualize electron-phonon coupling. Measurements, performed in the AP-mode configuration, were used to determine the overall UED temporal resolution in a polycrystalline sample. The time-stamping feature available in P-mode operation was not necessary for this calculation. A Bi thin film [Methods] was photoexcited with a 25 fs r.m.s. laser pulse ($\lambda$ = 800 nm, diameter of ~500 µm) at a pump fluence of ~0.19 mJ cm$^{-2}$. The laser pump transfers energy to the Bi lattice motion through the transient Debye–Waller-effect, causing atomic displacement perpendicular to its $c$-axis [35,36] on a time scale of $\tau_{Bi}$, dependent on the laser pump fluence.[35] The diffracted electrons were recorded as a function of the time delay between the laser pump pulse and electron probe pulse collected in a time window of 2.2 ps with a variable time step. Diffraction data were collected over multiple scans to ensure high signal-to-noise ratio, with drift and fast transition effects accounted for in post-processing [Methods]. Figure 3(a-b) shows the temporal evolution of the normalized diffraction intensity of the photoexcited Bi (300) ring. We performed a deconvolution process to obtain the intrinsic response time of Bi from the measured data to accurately determine the overall temporal resolution [Methods]. Under these operating conditions, the overall temporal resolution of our UED facility was 85 ± 28 fs r.m.s, compared to 178 ± 57 fs r.m.s. without this THz compression interaction, corresponding to about a factor of two in improvement. Critically, the achievable improvement is an intrinsic function of the initial electron probe length and jitter (in this case measured using the AP-mode mode to be 155 and 73 fs r.m.s. respectively).

**THE TEMPORAL RESOLUTION OF THE THZ TIME-STAMPED UED IN CRYSTALLINE AU FILMS**

A key advantage of this THz-compression scheme in the P-mode is that individual electron probe shots are time-stamped with a correlated time and position distribution. Here, we demonstrate the utilization of this technique to improve the temporal resolution of single-shot UED data by examining the response time and the transient dynamics of pump-probe measurements in a thin Au(100) film. A vertically polarized broadband THz pulse (central frequency 0.7 THz and FWHM of 0.45 THz, Fig. 4(b-c)) with peak electric field 20 MV/m producing a fluence of 50 µJ/cm$^2$ is used to trigger plasmonic resonances in the 11 nm Au film on Au TEM grid. Simulations of the THz interaction with the sample show transient oscillations occurring at frequencies beyond 1 THz as seen from the spectrum of the electric field in Fig. 4(c) (Supplementary Section 7). These features can only be observed by an electron probe with temporal length σ <100 fs r.m.s. and we use the UED measurement employing the time-stamping technique to resolve those details.

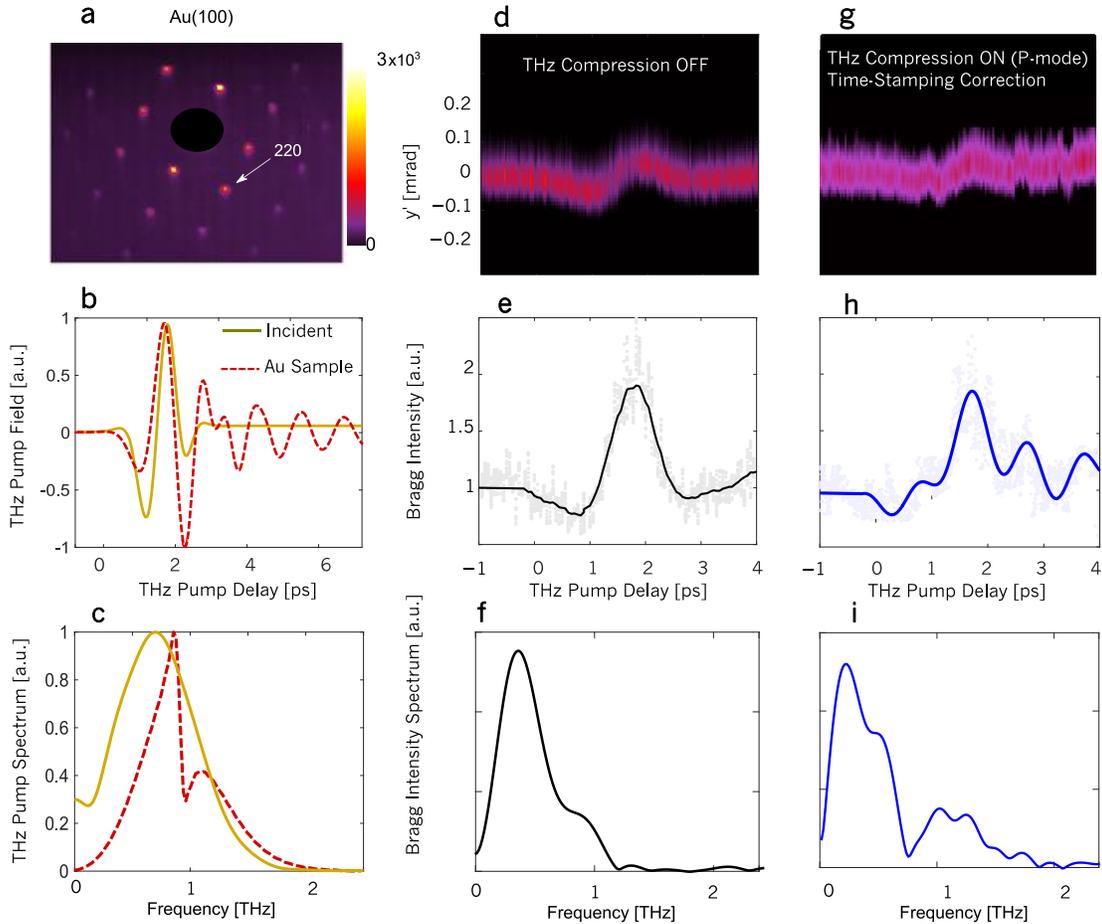

**Fig. 4 | Temporal resolution of THz compressed and time-stamped single shot UED probes with single-crystal Au sample in the P-mode.** (a) Diffraction pattern of Au(100) sample obtained from integrating 1000 shots, (b) THz pump field waveform generated from a LN setup and the corresponding waveform simulated based on interaction with the sample, (c) the spectrum of these THz waveforms in (b). (d) Measured time-dependent sliced beam transverse deflection along the y-axis of the (220) Bragg Peak of the single-crystal Au film as a function of the THz pump delay without THz compression, and (g) with THz compression and time-stamping correction showing a faster response time. (e-h) The corresponding normalized integrated intensity of the Bragg peak on a single shot basis (symbols) and averaged (lines), and (f-i) the Fourier transform of the integrated intensities in (g-h) with the average subtracted, showing the transient oscillations in the dynamical Bragg intensities for THz compression with time-stamping correction around 1.1 THz.

The time-dependent integrated intensity variations of the (200) Bragg peak in Au is correlated to the THz-induced transverse momentum of the electrons.[37] We show a comparison in the achieved temporal resolution between two cases in Fig. 4, with no THz compression, and with THz compression aided by time-stamping correction. In the latter, the time-stamp from the THz compression interaction is used *a posteriori* to correct the TOA jitter at the sample, using the algorithm in [Methods]. The time-dependent (220) Bragg spot transverse deflection (along the y-axis) in addition to the normalized integrated intensity due to the THz pump is depicted in Fig. 4(d,e,g,h), obtained on a single-shot basis over 5 ps. The normalized amplitude spectra through the Fourier transform of the integrated intensities with average subtracted are also shown Fig. 4(f,i). The overall temporal resolution, with no THz compression and with THz compression, was 181 fs and 56 fs r.m.s. respectively, improving the UED timing resolution by a factor of 3.2.

Furthermore, the Bragg peak intensities exhibit fast oscillations seen in the intensity spectrum when time-stamping correction is applied. We attribute these oscillations to resonances in the gold sample at frequencies >1 THz, which cannot be resolved as evident from the uncompressed case in Fig. 4(d-f).

Finally, to better visualize the single-shot time-stamping dynamics, we show the sliced Bragg peak intensity variations as a function of the THz delay. Each pixel slice of the time-stamped beam shows a similar transient response to the integrated beam intensity but exhibits a delay corresponding to the time-of-arrival of each slice with respect to the THz pump (Fig. 5(a-b)). The time-stamping resolution, defined as the linear slope of the time-of-arrival as a function of transverse position on the detector, is 5 fs/pixel or ~0.43 fs μrad$^{-1}$ accounting for the geometrical factor of our experiment. This means that the minimum temporal resolution that we are measuring on a single pixel of the detector is 5 fs. In Supplementary Fig. S7 we independently validate with the THz streaking diagnostic the time-stamping of the electron probe with measurement of the time-of-arrival of each beam slice. Similar results were also obtained in a thin Si film (Supplementary Section 6).

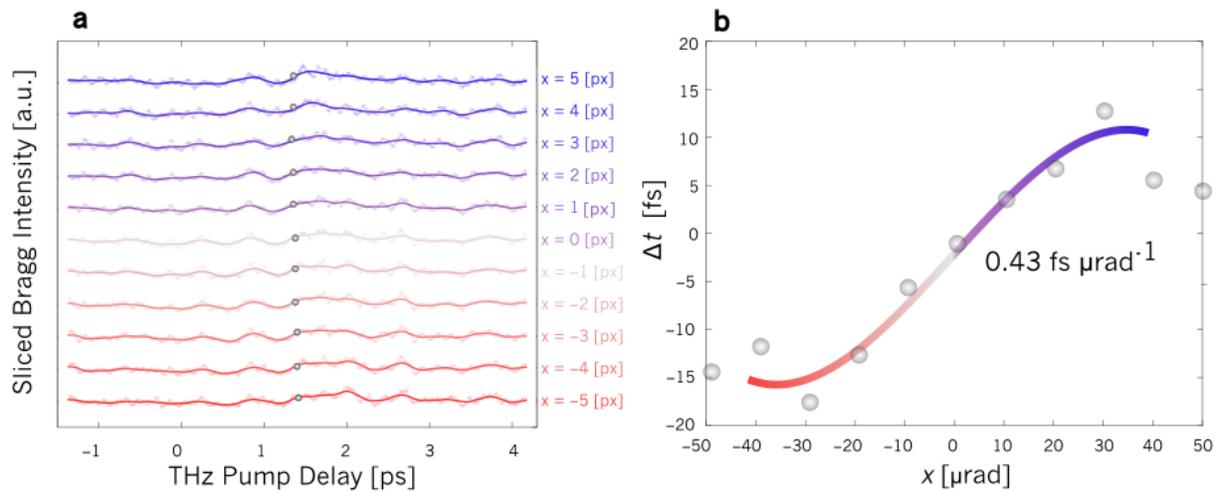

**Fig. 5 | Time-stamping of single-shot UED probes in THz-excited single-crystal Au in the P-mode.** (a) Stacked sliced Bragg spot intensities are taken on a per-pixel basis of the transverse beam distribution along *x*, with variation as a function of the THz pump delay. (b) The location of the maximum upward slope of the sliced intensity is designated by the marker and shown as a function of sliced angular location showing a resolution of ~ 0.43 fs·μrad$^{-1}$ in the linear region |x| <20 μrad, beyond which the beam exhibits a nonlinear pile-up from the THz compression interaction. Note that intensity data beyond the Bragg peak spot size of ~ 48 μrad exhibits high fluctuations due to reduced counts.

Note that in Fig. 5(b) the time-stamping exhibits nonlinearity form owing to the pile-up from the long initial electron probe length during the compressor interaction. Indeed, the linear proration of the single-cycle THz field in the compressor structure only extends up to 170 fs providing a linear time-stamping within ±73 μrad. Finally, we stress that the minimum temporal resolution that can be realized from this time-stamping scheme (in this work 5 fs, see Fig. 5(b)) is limited by the spatial resolution of the detector as well as the overall THz energy in the compressor stage (Supplementary Section 6).

In summary, we have demonstrated a new regime for UED utilizing THz compression to produce ultrashort electron probes with reduced TOA jitter. These electron probes were used to demonstrate improved temporal resolution as well as UED diffraction dynamics in crystalline materials. We demonstrated the THz dual-fed compression scheme in both P- and AP-modes, along with time-stamping in the P-mode. With THz-compression, the temporal resolution improvement of at least a factor of 2.5 to 85 fs is demonstrated through time-resolved diffraction measurements. Correcting for the mean electron probe arrival time improves the temporal resolution by a factor 3.2 to 56 fs. The ultimate resolution of the single-shot time-stamping measurement was demonstrated in single crystal samples to be around 5 fs. Furthermore, this technique is amendable to achieving better temporal resolution by further optimizing the THz pulse shape and preconditioning of the UV drive laser to produce shorter initial probes. Our technique, validated by these experimental measurements, indicates that given an initial 50 fs r.m.s. UED probe, this dual-fed THz compressor and time-stamping technique could produce an ultra-short electron probe sufficient for sub-fs temporal resolution at the downstream sample.

Detailed study of THz-pump-UED probe[37] in single-crystal samples can provide a further understanding of carrier dynamics and plasmon interactions,[37] as well as coherent phonon oscillations,[16,38] in previously inaccessible fs time scales. The time-stamping technique demonstrated here can also be used without THz compression nor THz pumping during normal UED operation with a pump laser. In this regime, a THz time-stamping stage can be positioned downstream from the diffraction detector to provide similar spatiotemporal correlation and jitter correction without influencing the UED measurements.

**ACKNOWLEDGMENTS**

The authors acknowledge M. Cardoso, and A. Haase for structure fabrication. This research has been supported by the U.S. Department of Energy (DOE) under Contract No. DE-AC02-76SF00515. The SLAC MeV-UED program is supported in part by the DOE Basic Energy Sciences (BES) Scientific User Facilities Division Accelerator & Detector R&D program, the Linac Coherent Light Source (LCLS) Facility, and SLAC under Contracts No. DE-AC02-05-CH11231 and No. DE-AC02-76SF00515. M. C. H. is supported by the DOE Office of Science, BES, Award No. 2015-SLAC100238-Funding. This work was also supported by NSF grant PHY-1734015.

**METHODS**

**Analysis of Beam Images**

Individual single shot beam image data collected from the electron-multiplying charged coupled devices (EMCCD) camera were analyzed and fitted to a two-Gaussian model. To measure the electron probe length and TOA from the THz streaking setup, we used a femtosecond per pixel conversion obtained by scanning the probe centroid as a function of the THz streaking delay. Our method also corrects the "pile-up" at the ends of the projected distribution and maps it to a position-dependent fs/pixel conversion. All images that did not pass a cutoff in the confidence interval of the probe length fit with the Gaussian distribution were removed from the final calculation, which only occurred in the original beam probes with no THz compression. An additional filter was applied that removed all data with a compressor time of arrival greater than ±100 fs from the minimum probe length and compression factor calculation. This reduced some of the effects of beam jitter in the result so that the minimum probe length value was only calculated from shots that had arrived at the compressor within a reasonable time frame. The standard deviation for the probe compression factor was calculated using standard error propagation.

**Determination of the UED Temporal Resolution from Pump-Probe Measurement**

The overall temporal resolution of the MeV UED facility can be determined as the r.m.s. of the contributing terms

$$\Delta t_{\text{overall}} = \sqrt{\Delta t_{\text{pump}}^2 + \Delta t_{\text{bunch}}^2 + \Delta t_{\text{jitter}}^2 + \Delta t_{\text{drift}}^2} \tag{1}$$

where $\Delta t_{\text{pump}}$ is the temporal pulse duration of the laser pump, $\Delta t_{\text{probe}}$ is the temporal pulse duration of the electron probe pulses, $\Delta t_{\text{jitter}}$ is the arrival time jitter between the pump and probe pulses at the sample position, $\Delta t_{\text{drift}}$ is the slow time drift. All values are used in r.m.s. convention. Other terms can also be added in (1) to account for velocity mismatches and jitter from the laser system which are ignored for simplicity.

The intrinsic response of Bi after photoexcitation with 800 nm laser is known and given by

$$I_{\text{Bi},0}(t) = \begin{cases} 1 - \delta\left(1 - \exp(-(t-t_0)/\tau_{\text{Bi}})\right), & t > t_0 \\ 1, & t < t_0 \end{cases} \tag{2}$$

where $\tau_{\text{Bi}}$ is the intrinsic response time-constant of Bi, and $\delta$ is the drop in diffraction signal following the pump. For the THz excited gold sample, we can also define $I_{\text{Au},0}(t) = \sum_n a_n \sin(\omega_n t)$ that takes into account the oscillations in diffraction intensity due to resonances. The UED instrument response is given by

$$H_{\text{UED}}(t) = H_0 \exp\left(\frac{(t-t_0)^2}{2\Delta t_{\text{overall}}^2}\right). \tag{3}$$

The measurement of the sample based on the actual instrument resolution is characterized by

$$I_{\text{sample}}(t) = \int_{-\infty}^{\infty} I_{\text{sample},0}(t') H_{\text{UED}}(t-t') dt'. \tag{4}$$

The estimated measured responses $I_{\text{Bi}}(t)$ and $I_{\text{Au}}(t)$ are then fitted to the measured diffraction signal using a least-square algorithm to determine the parameters.

UED diffraction patterns from photoexcited Bi film were collected in multiple scans to guarantee that the signal-to-noise ratio is sufficient for analysis. Here, the pump fluence is relatively low and the intrinsic response time is about 257±50 fs rms.[36] About 15 scans were performed with varying step size from 66 fs down to 33 fs and then the azimuthal average for each image was calculated. For each delay point, 5 to 10 images were collected each of which is an average of 36 single shots. Note that the P-mode THz compression is not ideal for polycrystalline samples as the transverse distribution of the beam may obscure some of the features in the ring diffraction patterns. For the THz-pumped Au sample UED resolution, more details are provided in Supplementary Section 6.

**Time-Stamping Algorithm**

The time-stamping correction algorithm utilized in this work uses the transverse beam (Bragg peak) distribution along the x-direction (in which the THz compressor induces spatiotemporal correlation) to correct for the time-of-arrival jitter of the probe with respect to the THz pump at the sample. For diffraction patterns obtained from a single crystal Au sample, beam images were taken at different stage positions, $t_s$, using a 50 fs step, with 20 images collected at each delay stage position. The actual time-of-arrival $t_{\text{TOA}}$ of every single shot can then be obtained by

$$t_{\text{TOA}} = t_s - D(x_s - x_r) \tag{5}$$

where $D$ is the calibration of the THz-induced time-stamping from the compressor interaction, $x_s$ is the transverse centroid of the beam and $x_r$ is the calibrated transverse centroid taken from a reference scan (with THz compression off). After evaluating $t_{\text{TOA}}$ for each shot and correcting for the jitter, we average shots with TOAs within ±20 fs to provide better statistics of beam distribution. Example beam distributions before and after time-stamping are reported in Supplementary S6. The beam intensity is then calculated per pixel in the image as in Fig. 5. In our model, we used $D$ = 4.5 fs/pixel.

**Sample Information**

The THz compressor structure was fabricated from OFE copper. The structure was then assembled from 6 individual parts to obtain a precise beam tunnel (radius of 45 μm). The minimum gap between the parallel plates is 75 μm. The structure is bolted together and mounted inside the vacuum chamber with a 3-axis motorized stage for coupling optimization of the THz pulses. The Bi sample is a 35 nm thin film grown on free-standing silicon nitride ($Si_3N_4$) membranes using molecular beam epitaxy (MBE) in the (100) orientation. The Au sample was obtained from Ted Pella, it is a large area single crystal gold film, approximately 11 nm thick grown in the (100) orientation and suspended on a 3 mm diameter gold TEM grid of thickness 20 µm. The TEM grid with the Au film supports transient oscillations at THz frequencies through THz interaction with the film and the periodic TEM grid (Supplementary Section 7).